\documentclass[a4,10pt,twocolumn]{article}
\usepackage{graphicx} 
\usepackage{mathptmx} 
\usepackage{times} 
\usepackage{amsmath} 
\usepackage{amssymb}  
\usepackage{amsthm}
\usepackage{tabularx}
\usepackage{float}
\usepackage[svgnames, table]{xcolor}    

\newtheorem{thm}{Theorem}[section]

\newtheorem{prop}[thm]{Proposition}

\newcommand{\todo}[1]{\textcolor{red}{ #1 }}
\renewcommand{\todo}[1]{}
\title{\LARGE \bf

A minimal dynamical system and analog \\circuit for non-associative learning
}
\author{
Matthew Smart$^1$, 
Stanislav Y. Shvartsman$^2$, 
and Martin M\"onnigmann$^3$
\footnote{
$^1$ Center for Computational Biology, Flatiron Institute {\tt msmart@flatironinstitute.org}; 
$^2$ Department of Molecular Biology and the Lewis-Sigler Institute of Integrative Genomics, Princeton University {\tt stas@princeton.edu}; 
$^3$ Ruhr University Bochum, Germany {\tt martin.moennigmann@rub.de}.
  }
}

\usepackage{etoolbox}
\makeatletter
\patchcmd{\maketitle}
 {\def\@makefnmark}
 {\def\@makefnmark{}\def\useless@macro}
 {}{}
\makeatother

\begin{document}
\maketitle
\thispagestyle{empty}
\pagestyle{empty}

\begin{abstract}
    Learning in \emph{living} organisms is typically associated with networks of neurons.
    The use of large numbers of adjustable units has also been a crucial factor in the continued success of \emph{artificial} neural networks. 
    In light of the complexity of both living and artificial neural networks, it is surprising to see that very simple organisms -- even unicellular organisms that do not possess a nervous system -- are capable of certain forms of learning. Since in these cases learning may be implemented with much simpler structures than neural networks, it is natural to ask how simple the building blocks required for basic forms of learning may be. The purpose of this study is to discuss the simplest dynamical systems that model a fundamental form of non-associative learning, habituation, and to elucidate technical implementations of such systems, which may be used to implement non-associative learning in neuromorphic computing and related applications. 
\end{abstract}

\section{INTRODUCTION}
\label{sec:Intro}
\typeout{col width is \the\columnwidth} 
A fascinating aspect of all living systems is their capacity to learn. 
One of the most basic forms of non-associative learning is habituation, which is characterized by a progressively diminishing response to repetitive stimulation.
This ability allows organisms, which are typically immersed in rich environments filled with benign, repetitive stimuli, to ``tune out" neutral signals and thus preserve limited cognitive resources and energy for responding to more salient ones (e.g. potential predators, prey, and mates). 
Despite its prevalence, dynamical mechanisms underlying habituation remain poorly understood. 

Although habituation is classically associated with animal (i.e. neuronal) learning, it has been observed across different kingdoms of life \cite{Gagliano2014, Boisseau2016, Boussard2019}. 
Thus, nervous systems are not a necessary ingredient for this fundamental form of learning. 
Remarkably, it has also been reported in various unicellular organisms \cite{Patterson1973, Eisenstein1982, Rajan2023} and in certain non-living systems~\cite{Zuo2017, Zhang2021resistance}.
In the latter works, which report the first technical implementations of habituation to the authors' knowledge, habituation of material conductance is mediated by repeated exposure to hydrogen gas, which transiently alters the electronic properties of the material (perovskite nickelate SmNiO$_3$~\cite{Zuo2017} or the Mott insulator nickel oxide~\cite{Zhang2021resistance}). 

Artificial implementations of non-associative learning are of practical interest because of their potential use in neuromorphic computing \cite{Zhang2020} and generalized physical learning \cite{Stern2023}.
In particular, Zuo et al.~\cite{Zuo2017} showed that a habituation-inspired extension of spiking artificial neural networks (ANNs) mitigates the phenomenon of catastrophic forgetting and 
outperforms the standard ANN in pattern recognition.
Similarly, Zhang et al.~\cite{Zhang2021resistance} demonstrated an artificial network inspired by non-associative learning has advantages for unsupervised clustering. 

In the present paper, we present a simple analog circuit that implements hallmarks of
non-associative learning, which essentially resulted from the authors' effort to find the simplest possible dynamical systems for habituation. 
Due to its simplicity, this circuit may ultimately help improve analog implementations of artificial learning systems, which are of interest, among other reasons, because they help close the gap in energy efficiency between living learning systems and their artificial counterparts (see, e.g., \cite{Zhang2020, LeGallo2023}). 

Non-associative learning is typically understood to comprise both habituation and sensitization, the progressive attenuation and amplification of a response to a repeated stimulus, respectively. 
We focus on habituation because it is the more complicated of the two processes 
(as will be obvious from the proposed circuit).
A seminal review by Thompson and Spencer~\cite{Thompson1966} outlined nine characteristic features associated with habituation in living systems. 
These features have since been revisited and refined~\cite{rankin2009} and are commonly known as the ``hallmarks of habituation". 
A starting point for our work is to restate these observational hallmarks in a mathematical setting (see Table~\ref{tab:hallmarks}). 
It is evident from their original statements that not all hallmarks need 
apply in order for an organism to be understood to habituate (cf.\ the ``may" in $H_6$ in Table~\ref{tab:hallmarks}, for example). 
In line with the common interpretation of the hallmarks, we consider the first sentence of hallmarks $H_1$ and $H_2$ (see Table~\ref{tab:hallmarks}) to be necessary conditions. 


\begin{table*}[ht]
    \noindent
    \caption{Hallmarks $H_1$, $H_2$, $H_6$, and $H_8$ from~\cite{Thompson1966, rankin2009} in shortened form, alongside proposed mathematical criteria.}
    \begin{tabularx}{\linewidth}{l X X}
        
        \hline
        $H_1$
        & 
        \begin{minipage}[t]{\linewidth}
        \cellcolor{Gray!5}
          \emph{Habituation}: 
          Repeated application of a stimulus results in a progressive decrease of a response to an asymptotic level. 
        \end{minipage}
        & 
        There exists a periodic stimulus $u(t)$ that generates a sequence of responses satisfying 
        $y[k+1] \le y[k]$ for all $k\ge 0$, and there exists a $K>0$ such that $y[k+1] < y[k]$ for all $k = 0, \ldots, K$. 
        \\

        \hline
        $H_2$
        & 
        \begin{minipage}[t]{\linewidth}
        \cellcolor{Gray!5}
          \emph{Spontaneous recovery}:
          If the stimulus is withheld, the response recovers over time.
        \end{minipage}
        & 
        Assume $H_1$ holds for $u(t)$. 
        Suppose the stimulus has been applied $k$ times resulting in a response $y[k]$. 
        Then there exists an $m \in \mathbb{N}$ such that the response after withholding the stimulus for $m$ periods satisfies $y[k+m+1] > y[k]$.
        \\

        \hline
        $H_6$
        & 
        \begin{minipage}[t]{\linewidth}
        \cellcolor{Gray!5}
          \emph{Subliminal accumulation}:
          The effects of repeated stimulation may continue to accumulate even after the response has reached an asymptotic level. Among other effects, this can delay the onset of spontaneous recovery.
        \end{minipage}
        & 
        Assume $H_1$ holds for $u(t)$. 
        Suppose the response reaches an asymptotic level on the $k^{th}$ stimulation. 
        Let $l_1 < l_2 \in \mathbb{N}$. 
        Then a system receiving $k+l_1$ stimulations will take longer to recover than a system receiving $k+l_2$ stimulations.
        \\
        
        \hline
        $H_8$
        & 
        \begin{minipage}[t]{\linewidth}
        \cellcolor{Gray!5}
          \emph{Dishabituation}:
          Presentation of another (usually strong) stimulus results in recovery of the habituated response.
        \end{minipage}
        & 
        Assume $H_1$ holds for $u(t)$. 
        There exists a second stimulus $s(t)$ applied only during $kT < t < (k+1)T$ which guarantees $y[k + 1] > y[k]$.
        \\
        \hline
        
    \end{tabularx}
    \label{tab:hallmarks}
\end{table*}

We pose the question of how simple a habituating dynamic system may be and answer this question from a systems \& control perspective. 
In the systems biology literature, this question would be paraphrased to read \emph{What does a motif for habituation look like?} (see, e.g., \cite{Alon2006} for a classical text on motifs and their nonlinear dynamics). 

An electrical circuit that mimics habituation in living systems was given in \cite{elife2023miami}. There, the circuit was linear, however, and a nonlinear post-processing of the signals generated by the circuit was required. The circuit proposed here generates the habituating response by itself and we show any circuit capable of this must be nonlinear.  

We present the proposed circuit and a corresponding dynamical system that respect hallmarks $H_1$ and $H_2$ in Section~\ref{sec:Circuit}, and show how the circuit can be extended to account for additional hallmarks.   
It will be evident that the circuit and system, due to their simplicity, are not amenable to further simplifications. 
We prove that the nonlinearity of the circuit cannot be removed in Section~\ref{sec:LTIsystemsDoNotHabituate}. 
Conclusions and an outlook are stated in Section~\ref{sec:ConclusionsOutlook}. 

\subsection*{Notation and Preliminaries}
All systems are single-input-single-output systems unless stated otherwise. 

Biological systems and their signals are bounded. 
In cellular signaling cascades, for example, signals are coded as concentrations of biochemical species, which implies they are bounded from below because concentrations are nonnegative. 
Similarly, membrane potentials of nerve cells cannot exceed lower and upper bounds in living organisms. 
We assume without restriction that signals are bounded from below by zero.
Furthermore, we assume without restriction the origin is a steady state that results in a zero output signal. 

We model periodic input stimuli by rectangular functions with an amplitude $A$, period $T$, and duty $d$. We abbreviate $u(kT)$ by $u[k]$ where convenient. Let $y[k]$ denote the largest output that results in period $k$, i.e., $ y[k]= \max \{ y(t) \vert t\in [kT, (k+1)T) \}$, which always exists for the systems considered here. 

\section{PROPOSED SYSTEM AND \\CIRCUIT}
\label{sec:Circuit}
\begin{figure}
    \centering
    \includegraphics[width=0.85\linewidth]{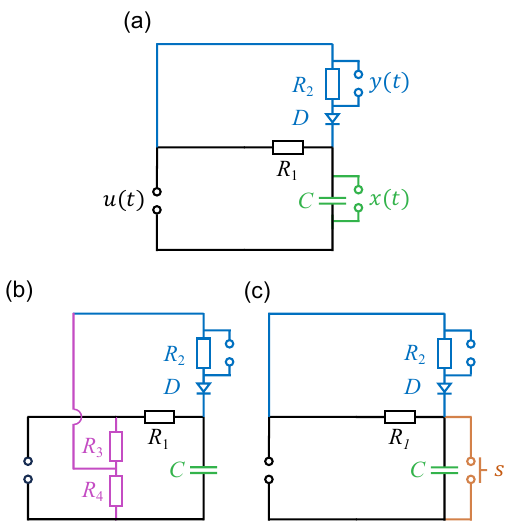}
    \caption{
    (a) Proposed circuit with input $u(t)$ and output $y(t)= U_{R_2}$, the voltage across $R_2$. 
    (b) Proposed circuit with an additional voltage divider. This extension is expected to be useful in future analyses of series and parallel connections of (a); analogous generalizations have been used before to cover additional hallmarks (see, e.g.~\cite{Staddon1993,eckert2022}). 
    (c) Extension by a path that discharges $C$ for $s=1$. The additional signal $s(t)$ is used to implement sensitization control for hallmark $H_8$ in Section~\ref{subsec:simulationResults}. 
    }
    \label{fig:CircuitDiagram}
\end{figure}

The proposed circuit and two useful extensions are shown in Figure~\ref{fig:CircuitDiagram}. 
Modeling this circuit is straightforward but results in a differential-algebraic model with an output that cannot be stated in explicit form due to the diode. 
The resulting model is stated in Section~\ref{sec:nonlinearCircuitModel}. 
We derive a simplified model in Section~\ref{sec:piecwiseAffineModel} that is still nonlinear but in explicit input-output form. 
It is shown in Section~\ref{subsec:simulationResults} that the proposed circuit habituates. 

\subsection{Nonlinear differential-algebraic model}\label{sec:nonlinearCircuitModel}
The proposed circuit is shown in Figure~\ref{fig:CircuitDiagram}a. 
Let $u(t)$, $x(t)$, and $y(t)$ be 
the input signal, 
the voltage $x = U_C$ across the capacitor, 
and the voltage $y= U_{R_2}$ across resistor $R_2$, 
respectively (see Figure~\ref{fig:CircuitDiagram}). 
Let $T_\text{RC}$ denote the time constant of the RC subcircuit in Figure~\ref{fig:CircuitDiagram}, i.e., $T_\text{RC}= R_1 C$. 
If $R_2\gg R_1$, the circuit can be modeled with
\begin{subequations}\label{eq:nonlinearCircuitModel}
\begin{align}  
    & T_\text{RC}\dot{x}+ x= u
    \label{eq:RCcircuit}
    \\
    & y= R_2 I_D \label{eq:nonlinearCircuitModelOutput}
\end{align} 
where the current $I_D$ through the diode and the voltage $U_D$ across the diode are given by the diode equation
\begin{equation}\label{eq:Diode}
    I_D= I_S \left(e^{\frac{U_{D}}{nU_T}}-1\right)
    \mbox{ and }
    U_D= \frac{R_4}{R_3+ R_4}\, u-x -R_2 I_D
\end{equation}
\end{subequations}
and where $c= 1$ 
for the circuit in Figure~\ref{fig:CircuitDiagram}a 
and $c= \frac{R_4}{R_3+ R_4}$ for the circuit in Figure~\ref{fig:CircuitDiagram}b.
Appendix~\ref{app:nonlinearCircuitModel} summarizes the derivation of~\eqref{eq:nonlinearCircuitModel}. 

\begin{figure}
    \centering
    \includegraphics[width=0.6\linewidth]{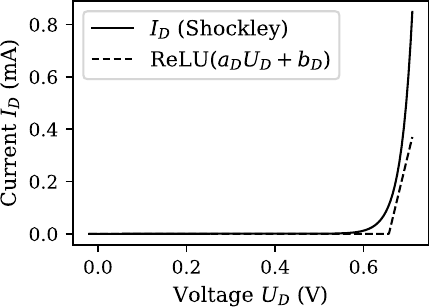}
    \caption{Diode equation $I(U_D)$ from~\eqref{eq:Diode} and piecewise linear approximation~\eqref{eq:piecewiseApproxShockley} with parameters from Table~\ref{tab:Parameters}.}
    \label{fig:PiecewiseApproxShockley}
\end{figure}

\subsection{Piecewise linear model}\label{sec:piecwiseAffineModel}
By approximating the diode equation~\eqref{eq:Diode} with
\begin{equation}\label{eq:piecewiseApproxShockley}
  I_D= \text{ReLU}\left(a_D U_D+ b_D\right),
\end{equation}
where $\text{ReLU}(z)= \max(0, z)$ is the rectified linear unit, 
the differential-algebraic model~\eqref{eq:nonlinearCircuitModel} 
can be simplified to
\begin{subequations}\label{eq:piecewiseAffineModel}
    \begin{align}
        & T_\text{RC} \dot{x} + x= u 
        \label{eq:piecewiseAffineModelState}
        \\
        & y= a\cdot \text{ReLU}(c u- x +b)\label{eq:piecewiseAffineModelOutput}
    \end{align}
\end{subequations}

The parameters in~\eqref{eq:piecewiseAffineModel} are given by $T_\text{RC}= R_1 C$ and $c$ already introduced in Section~\ref{sec:nonlinearCircuitModel} and
\begin{equation}\label{eq:piecewiseAffineModelConstants}
    a= \frac{a_D R_2}{1+ a_D R_2}, 
    \quad
    b= \frac{b_D}{a_D}.
\end{equation}
Note that, in contrast to \eqref{eq:nonlinearCircuitModel}, no algebraic equations need to be solved in~\eqref{eq:piecewiseAffineModel} to determine the output, which results from evaluating~\eqref{eq:piecewiseAffineModelOutput}.

Appendix~\ref{app:piecewiseAffineModel} briefly explains how to derive~\eqref{eq:piecewiseAffineModel}.

\subsection{The proposed circuit and system \\habituate}\label{subsec:simulationResults}
We first illustrate the approximation~\eqref{eq:piecewiseApproxShockley} of the diode equation~\eqref{eq:Diode} in Figure~\ref{fig:PiecewiseApproxShockley} for completeness. Values for $a_D$ and $b_D$, which are given in Table~\ref{tab:Parameters}, have been identified manually for the output from the explicit model~\eqref{eq:piecewiseAffineModel} to closely resemble the output from the original model~\eqref{eq:nonlinearCircuitModel}.

\begin{figure}[h]
    \centering
    \includegraphics[width=0.85\linewidth]{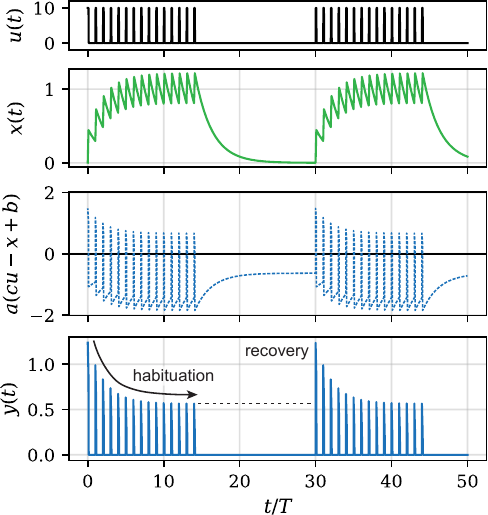}
    \caption{
        System fulfills $H_1$ and $H_2$ (habituation and spontaneous recovery, respectively; see Table~\ref{tab:hallmarks}). 
        As explained in the text, the RC subcircuit implements a leaky memory $x(t)$ of recent inputs $u(t)$ to the system, which is then compared with the input signal to generate the response $y(t)$. Notation: $u(t)$ denotes the input voltage, $x(t)$ denotes the voltage across the capacitor, and $y(t)$ denotes the voltage across resistor $R_2$ (all in $\mathrm{V}$).
        The third plot shows the (scaled) argument of the nonlinearity in 
        (\ref{eq:piecewiseAffineModelOutput}) as a dashed line.
        }
        
    \label{fig:simulationForH1andH2}
\end{figure}

It is evident from Figure~\ref{fig:simulationForH1andH2} that the circuit, more specifically the differential-algebraic model~\eqref{eq:nonlinearCircuitModel}, respects hallmarks 1 and 2. The RC circuit subsystem integrates the rectangular stimuli from its initial condition $x(0)= 0$ until a forced limit cycle is reached for $x(t)$ (green curve in Figure~\ref{fig:simulationForH1andH2}). 
The first order lag implemented by the RC circuit and its state $x$ essentially implement a memory variable that, roughly speaking, counts sufficiently fast stimuli (e.g. green curve for $0\le T\le 15T$ in Figure~\ref{fig:simulationForH1andH2}) and forgets these stimuli over time (e.g. green curve for $16T\le t\le 30T$ in Figure~\ref{fig:simulationForH1andH2}).

The output essentially subtracts the memory variable from the input signal (up to scaling with $a$ and $c$ and a constant shift $b$), 
which is evident from the circuit and the argument $cu-x+b$ of the ReLU in~\eqref{eq:piecewiseAffineModelOutput}. 
Due to the diode, negative voltages across $R_4$ are not passed to the output, which is again evident from the ReLU in~\eqref{eq:piecewiseAffineModelOutput} and $I_D>0$ in the diode equation in~\eqref{eq:Diode}.

As a result of these two operations, subtraction and rectification, the output $y(t)$ shows the desired behavior for hallmark $H_1$ in response to periodic stimuli (blue curve in Figure~\ref{fig:simulationForH1andH2}, up to $t= 15T$). When the stimuli pause (blue curve, $16T\le t< 30T$), the circuit forgets about the earlier stimuli ($x(t)$ approaches zero around $t= 30T$). The first subsequent stimulus ($t= 30T$) results in the same response as the stimulus at $t= 0$, which implies the circuit recovered spontaneously, i.e. it respects hallmark $H_2$. 

\begin{figure}
    \centering
    \includegraphics[width=0.85\linewidth]{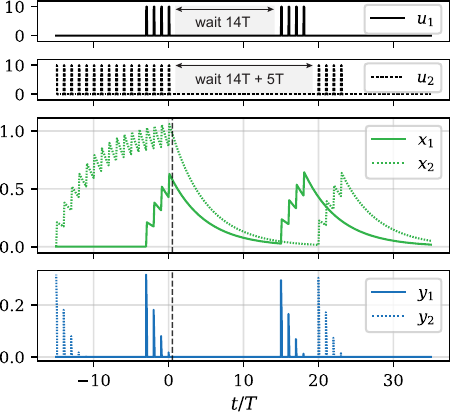}
    \caption{
        The proposed circuit and system fulfill hallmark $H_6$, subliminal accumulation, if the components are chosen to achieve an asymptotic value zero for $y(t)$ when subjected to periodic stimuli. 
        Solid curves $(u_1, x_1, y_1)$: Stimulation with a comb of length $4$ is sufficient to approach the asymptotic output value $y=0$. The system requires a delay of $14$ periods after the first comb ends to recover nearly fully when stimulated again at $t=15T$. 
        Dashed curves $(u_2, x_2, y_2)$: When stimulated with a longer first comb, the memory variable $x(t)$ can reach higher values. As a result, the system recovers later, about $19$ periods after the first comb ends.  
        In both simulations (solid and dashed), the first combs are aligned to stop at $t=0$ to facilitate comparison.  
        Distinct parameters for these simulations relative to Table~\ref{tab:Parameters} are 
        $R_1 = 1 \,\mathrm{k \Omega}$,
        $R_3 = 18 \,\mathrm{k \Omega}$,
        $R_4 = 2 \,\mathrm{k \Omega}$.
        }
    \label{fig:simulationForH6}
\end{figure}
Figure~\ref{fig:simulationForH6} shows that the system respects hallmark $H_6$ for some values of its components. These values are altered for the simulations shown in Figure~\ref{fig:simulationForH6} for better visibility of the desired effect. We claim without giving details that all behaviors shown in all figures are robust with respect to the component values, i.e. they do not appear for singular choices of the components. 
In the case shown in Figure~\ref{fig:simulationForH6}, after four stimulations the system reaches an asymptotic output value of $y=0$ (which it cannot exceed due to the $\mathrm{ReLU}$-like nonlinearity of the diode). 
If the system continues to be stimulated after reaching this asymptotic response 
(dashed curves in Figure~\ref{fig:simulationForH6}), 
it exhibits subliminal accumulation (hallmark $H_6$). 
Prior to the second round of stimulation, this effect is not directly visible from the response $y$ alone (hence \emph{subliminal} as in the original biological literature). The effect is transparent if one has access to the memory variable $x(t)$: the overstimulated system requires a longer duration to recover fully ($20T$ instead of $15T$) due to the accumulation of $x_2$ relative to $x_1$. 
This effect can be made more or less pronounced (or even removed) depending on the parameters of the circuit elements.

\begin{figure}
    \centering
    \includegraphics[width=0.85\linewidth]{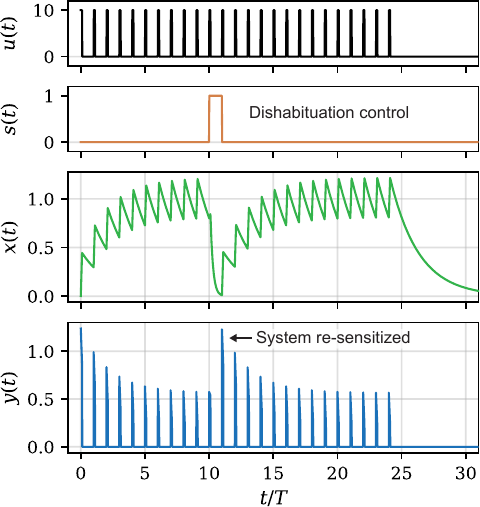}
    \caption{System fulfills $H_8$ (dishabituation; see Table~\ref{tab:hallmarks}). This re-sensitization is physically implemented using a switchable signal $s(t)$ that discharges the memory stored in the capacitor (see   Figure~\ref{fig:CircuitDiagram}C).}
    \label{fig:simulationForH8}
\end{figure}

Finally, Figure~\ref{fig:simulationForH8} shows that $s(t)$ introduced in Figure~\ref{fig:CircuitDiagram}c achieves sensitization (or equivalently,   \emph{dishabituation} to use the terminology of hallmark $H_8$). Closing the switch at time $t= 10T$ results in discharging the capacitor to $U_C= x= 0$, which is equivalent to resetting the RC circuit to its initial discharged state or forcing the memory variable to forget all earlier stimuli. In the simulation shown in Figure~\ref{fig:simulationForH8}, the switch had a resistance of $47\Omega$ in its closed state which gives rise to an exponential discharge curve for $t\in [10T, 11T)$ in the figure.    


\begin{table}[h]
    \label{tab:Parameters}
    \begin{center}
        \begin{tabular}{ll}
            \hline
            $R_1= 470 \,\Omega$ & $R_2= 10 \,\mathrm{k}\Omega$ \\
            $R_3= 8.2 \,\mathrm{k}\Omega$ & $R_4= 2.2 \,\mathrm{k}\Omega$ \\
            $I_S= 1\cdot 10^{-15} \,\mathrm{A}$ & $n=1$ \\
            $U_T= 25.85\,\mathrm{mV}$ & $C = 4700\,\mathrm{\mu F}$ \\
            $a_D= 7.0\,\mathrm{mS}$ & $b_D= -4.6\,\mathrm{mA}$ \\
            
            \hline
            \end{tabular}
        \end{center}
    \caption{
        Parameters used in the circuit simulations. Voltages shown in Fig. \ref{fig:simulationForH1andH2}-\ref{fig:simulationForH8} are in $\mathrm{V}$ unless specified otherwise. For the input stimulus $u(t)$ we use $T=1\,\mathrm{s}$ and $d=0.1$ throughout. 
    }
\end{table}

\section{LTI SYSTEMS DO NOT \\HABITUATE}\label{sec:LTIsystemsDoNotHabituate}
It is instructive to consider if the proposed circuit and system can be simplified further.
Since the diode $D$ is the only nonlinear component in the proposed circuit, it is natural to ask whether the diode can be omitted, or more generally, whether there exists a \emph{linear} system that habituates. 

We show linear time-invariant systems cannot habituate in this section. 
We use a behavioral setting and notation (see, e.g.~\cite{Sontag1998,PoldermanWillems1998}) in order to not restrict the result to ordinary differential equations or their discrete-time counterparts. 

While the claim and proof stated in Proposition~\ref{prop:LTIsystemsDoNotHabituate} are a bit technical due to their intended generality, the essential idea is simple and sketched in Figure~\ref{fig:LTISystemsDoNotHabituate}: Assuming a system habituates in response to a signal $u$, it habituates in the same fashion if $u$ is shifted to a later time (due to time invariance, Figure~\ref{fig:LTISystemsDoNotHabituate}a and b). Linearity implies that the response to $u$ can be superimposed with the response to the time-shifted $u$ to obtain the response to the combined inputs (Figure~\ref{fig:LTISystemsDoNotHabituate}c). However, if the system habituates, more specifically respects hallmark $H_1$, it responds differently to only the time-shifted $u$ than to the superposition of $u$ and the time-shifted $u$ (Figure~\ref{fig:LTISystemsDoNotHabituate}c and d differ). 


\todo{Our system has an intrinsic time scale, here $T_\text{RC}$, that determines which range of signals the system habituates to. Is there a system that learns this timescale?}

As a preparation to Proposition~\ref{prop:LTIsystemsDoNotHabituate}, 
let a dynamical system be defined by 
\begin{equation}\label{eq:DynamicalSystemTupel}
    \left(\mathcal{T}, \mathcal{U}, \mathcal{Y}, \mathfrak{B}\right)
\end{equation}
where $\mathcal{T}\subseteq\mathbb{R}$ is a set of time points and 
$\mathcal{U}$ and $\mathcal{Y}$ are sets of functions on $\mathcal{T}$ that describe stimulus and responses, respectively. $\mathfrak{B}$ is the subset of all functions from $\mathcal{T}$ to $(\mathcal{U}\times\mathcal{Y})$ that describe the behavior of the system, i.e., $(u\times y)\in \mathcal{U}\times\mathcal{Y}$ implies the system can react with the response $y(t)$, $t\in\mathcal{T}$ to the stimulus $u(t)$, $t\in\mathcal{T}$.  
A dynamical system~\eqref{eq:DynamicalSystemTupel} is said to be \emph{linear}, 
if, for any
$(u_1, y_1)\in\mathfrak{B}$, $(u_2, y_2)\in\mathfrak{B}$ and any $c_1\in\mathbb{R}$, $c_2\in\mathbb{R}$, we have
\begin{equation}\label{eq:Superposition}
    (c_1 u_1+ c_2 u_2, c_1 y_1+ c_2 y_2)\in\mathfrak{B} .
\end{equation}
A dynamical system~\eqref{eq:DynamicalSystemTupel} is said to be \emph{time-invariant}, if, for any $t^\prime\in\mathcal{T}$ and any $(u, y)\in\mathfrak{B}$, we have
$(u^\prime, y^\prime)\in\mathfrak{B}$, where $u^\prime$ is defined by $u^\prime(t)=u(t-t^\prime)$ and $y^\prime(t)$ is defined accordingly. 

\begin{prop}\label{prop:LTIsystemsDoNotHabituate}
    A linear time-invariant system with a nonnegative output cannot respect hallmark $H_1$ and thus cannot habituate. 
\end{prop}
\begin{proof}
We assume there exists a system with the stated properties that respects hallmark $H_1$ and show that a contradiction results. 
We assume without restriction that the system state is the origin and $y=0$ before the first stimulus applies.

Let $u$ be a sequence of $L$ stimuli with period $T$, where the first stimulus occurs at $t= 0$.  
Let $y$ be such that $(u, y)\in\mathfrak{B}$. 
Define $u^\prime$ by $u^\prime(t)= u(t-LT)$ for all $t\in\mathcal{T}$, define $y^\prime$ accordingly, and note that this implies $y^\prime[k+L]= y[k]$ for all $k\in\mathbb{N}$. 
Since the system is time-invariant, 
$(u, y)\in\mathfrak{B}$ implies 
$(u^\prime, y^\prime)\in\mathfrak{B}$. 
%
Since the system fulfills hallmark $H_1$ by assumption, we have $y[0]> y[k]$ for all $k\in\mathbb{N}$, since there exists a $K>0$ such that $y[k]$, $k= 0, \dots, K$ is strictly decreasing. This implies and $y^\prime[L]> y^\prime[L+k]$ for all $k\in \mathbb{N}$, since $y^\prime[L+ k]= y[k]$ for all $k\in\mathbb{N}$.

\begin{figure}
    \centering\includegraphics[width=0.9\linewidth]{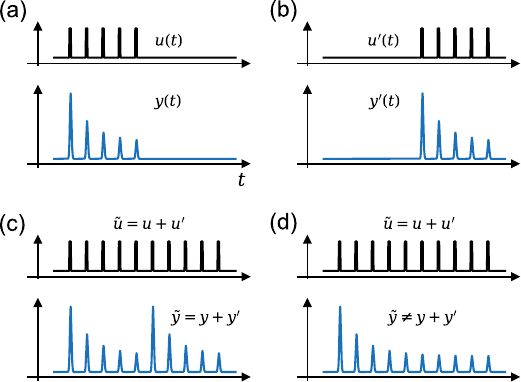}
    \caption{Linear time-invariant systems do not respect hallmark $H_1$. Linear superposition according to~\eqref{eq:Superposition} of $(u,y)$ shown in (a) and $(u^\prime, y^\prime)$ shown in (b) results in the time series shown in (c). 
    The response in (c) does not respect hallmark $H_1$, however, in contrast to the response in (d).
    }
    \label{fig:LTISystemsDoNotHabituate}
\end{figure}

Now consider $\tilde{u}= u+ u^\prime$ and $\tilde{y}= y+ y^\prime$, 
and 
note that $(\tilde{u}, \tilde{y})\in\mathcal{B}$ since the system is linear.  
Since the system respects hallmark $H_1$ by assumption, we have 
\begin{equation}\label{eq:LinSysViolateHallmark1Helper1}
    \tilde{y}[L]< \tilde{y}[0]= y[0]+ y^\prime[0]= y[0]
\end{equation}
where we used $y^\prime[0]= 0$, which holds because no stimulus has appeared for $(u^\prime, y^\prime)$ at time $t= 0$. 
Since the system is linear by assumption, we have
\begin{equation}\label{eq:LinSysViolateHallmark1Helper2}
    \tilde{y}[L]= y[L]+ y^\prime[L]= y[L]+ y[0]\ge y[0]
\end{equation}
where the first equality holds due to linearity, the second one due to time invariance and the last relation holds with $y[L]\ge 0$, i.e. nonnegativity of the output.
Relations~\eqref{eq:LinSysViolateHallmark1Helper1} and~\eqref{eq:LinSysViolateHallmark1Helper2} yield 
$\tilde{y}[L]< y[0]$ and $\tilde{y}[L]\ge y[0]$, which is the desired contradiction.
\end{proof}
\todo{Point out that the system class includes continuous-time and discrete-time systems, systems governed by ordinary or partial differential equations, maps, NNs, ... .}

We pointed out in Section~\ref{sec:Intro} that biological systems are bounded. 
Since linear systems and their outputs are \emph{unbounded} according to the superposition principle, it is tempting to infer that linear systems cannot habituate from the assumption of bounded outputs alone. Input signals to living systems are, however, also bounded, and properties like bounded-input-bounded-output stability suggest bounded outputs can be ensured under certain assumptions on the input signals. As a result, the mere statement that linear systems are ruled out here due to their unboundedness is arguably not satisfactory. Proposition~\ref{prop:LTIsystemsDoNotHabituate} and its proof make the statement clearer. Note that boundedness enters in~\eqref{eq:LinSysViolateHallmark1Helper2} in the proof.

\section{CONCLUSIONS AND \\OUTLOOK}\label{sec:ConclusionsOutlook}

We have described a minimal analog circuit that implements a fundamental form of non-associative learning, habituation, in that it satisfies its two defining features (or \emph{hallmarks}, specifically $H_1$ and $H_2$ in Table~\ref{tab:hallmarks}). 
We have shown that additional features of habituation found in living organisms can be achieved with mild extensions to the base circuit.  

The simplicity of the proposed circuit 
may facilitate applications in 
neuromorphic computing~\cite{Zhang2020} or, more generally, physical learning~\cite{Stern2023}. It has recently been shown that non-neuronal learning can provide benefits relative to conventional
ANNs~\cite{Zhang2021resistance}. 
Artificial habituation system have been proposed for this purpose~\cite{Zuo2017,Zhang2021resistance} that are based on changing the electronic properties of nickelates or nickel oxide by exposing them to hydrogen and ozone. 
Due to its flexibility, the circuit proposed here may be an interesting alternative. 
For instance, the circuit can readily be adjusted for a wide range of timescales ($T_\text{RC}$ in~\eqref{eq:RCcircuit}, \eqref{eq:piecewiseAffineModelState}). 

Future work will address extensions of the base circuit that account for remaining hallmarks of habituation. 
In particular, series connections of more complex but related models have been suggested to exhibit frequency sensitivity \cite{eckert2022, Staddon1996}. 
Thus, a natural next step is to treat the presented circuit as a building block and investigate the properties of series and parallel assemblies. Such architectures will allow to extend beyond the single-input, single-output setting to implement more elaborate sensory and response behaviors.

\section*{ACKNOWLEDGMENTS}
    We thank Anton Persikov, Hayden Nunley, and Gilles Francfort for helpful discussions and suggestions. M.M.\ gratefully acknowledges travel support by the Simons Foundation. M.S.\ and S.Y.S.\
    are grateful for ongoing support through the Flatiron Institute, a division of the Simons Foundation.



\appendix

\section{Derivation of~\eqref{eq:nonlinearCircuitModel}
}\label{app:nonlinearCircuitModel}
Resistors $R_3$ and $R_4$ form a voltage divider that provides the potential 
\begin{equation}\label{eq:UR2old}
    U_{R_4}= \underbrace{R_4/(R_3+ R_4)}_{c}u(t).
\end{equation}
If $R_2\gg R_1$, the RC circuit formed by $R_1$ and $C$ (black subcircuit in Figure~\ref{fig:CircuitDiagram}) is approximately independent of the remaining components. 
The voltage across $C$, $x= U_{C}$, can therefore be modeled with~\eqref{eq:RCcircuit} with time constant $T_\text{RC}= R_1 C$.
The potential $U_{R_4}$ from~\eqref{eq:UR2old} applies to the series connection of $R_2$, $D$ and $C$ (highlighted in purple), which implies
\begin{equation}\label{eq:Kirchhoff}
    U_{R_4}= U_{R_2}+ U_{D}+ U_{C},
\end{equation}
where $U_D$ and $U_{C}$ are the voltages across $D$ and $C$, respectively. 
The current $I_D$ through $D$ can be modeled with Shockley's diode equation stated in~\eqref{eq:Diode},
where $I_S$, $n$, and $U_T$ are the scale current of the diode, the ideality factor of the diode, and the thermal voltage, respectively.
Since the currents through $R_2$ and $D$ are equal, 
\begin{equation}\label{eq:UR4old}
    U_{R_2}= R_2 I_D.
\end{equation}
Substituting~\eqref{eq:UR2old}, \eqref{eq:UR4old} and $U_{C}= x$ into~\eqref{eq:Kirchhoff} and rearranging yields the second equation in~\eqref{eq:Diode}. 

\section{Derivation of~\eqref{eq:piecewiseAffineModel}}\label{app:piecewiseAffineModel}
Assume $a_D U_D+ b_D\ge 0$ which implies the approximated diode equation~\eqref{eq:piecewiseApproxShockley} evaluates to $I_D= a_D U_D+ b_D$. 
The explicit output equation~\eqref{eq:piecewiseAffineModelOutput} can be found by substituting $I_D= a_D U_D+ b_D$ into~\eqref{eq:UR4old}, substituting the resulting expression into~\eqref{eq:UR2old} and solving for $U_{D}$. 
This yields
\begin{align*}
    U_{D} &= \frac{1}{1+a_D R_2}\left(U_{R_4}- U_{C}- b_D R_2\right)\\
     &= \frac{1}{1+a_D R_2}\left(c u- x- b_D R_2\right),
\end{align*}
where~\eqref{eq:UR2old} and $x= U_{C}$ were used in the last step. 
Substituting this expression for $U_{D}$ into $I_D= a_D U_D + b_D$ and substituting the resulting expression into~\eqref{eq:UR4old} yields
\begin{align}
    U_{R_2}&= \frac{a_D R_2}{1+a_D R_2}\left(c u- x- b_D R_2\right)+ b_D R_2
    \nonumber\\
    &= \frac{a_D R_2}{1+ a_D R_2}\left(cu- x- b_D R_2\left(1-\frac{1+a_D R_2}{a_D R_2}\right)\right)
    \nonumber\\
    &= \frac{a_D R_2}{1+ a_D R_2}\left(cu- x + \frac{b_D}{a_D}\right) 
    \label{eq:UR4oldhelper1}
\end{align}
This expression was derived assuming $I_D= a_D U_D + b_D$, i.e., for one branch of the approximate diode equation~\eqref{eq:piecewiseApproxShockley}.
The other branch, $I_D= 0$, implies 
\begin{equation}\label{eq:UR4oldhelper2}
    U_{R_2}= R_2 I_D= 0.
\end{equation} 
In summary, $y= U_{R_2}$ is given by~\eqref{eq:UR4oldhelper1} if $a_D U_D+ b_D\ge 0$ and given by~\eqref{eq:UR4oldhelper2} if $a_D U_D+ b_D< 0$, which is equivalent to ~\eqref{eq:piecewiseAffineModelOutput} with $a$ and $b$ defined in~\eqref{eq:piecewiseAffineModelConstants}. 
\newline


\bibliography{references}
\bibliographystyle{abbrv}
\end{document}